%% file: long_version.tex
\documentclass[conference,letterpaper]{IEEEtran}
\input{latex_def.tex}
\def \LONG{}
\begin{document}
\input{main.tex}

\bibliographystyle{IEEEtran}
\bibliography{clique}

\appendices
\crefalias{section}{appsec}

\section{Proofs for \cref{sec:p-rem}}
\label{app:rem}
\input{planted_rem_appendix.tex}

\section{Proofs for \cref{sec:w-sbm}}
\label{app:k_subgraph}
\input{planted_k_subgraph_appendix.tex}

\section{Proof for \cref{sec:w-hsbm}}
\label{app:hypergraph}
\input{appendix_w_hsbm.tex}

\end{document}

%% file: latex_def.tex
\usepackage[utf8]{inputenc} 
\usepackage[T1]{fontenc}
\usepackage{url}
\usepackage{ifthen}
\usepackage{cite}
\usepackage[cmex10]{amsmath} 
\usepackage{amsfonts,mathtools}
\usepackage{bbm}
\usepackage{cleveref}
\usepackage{color}
\usepackage{amsmath,amssymb}

\DeclareMathOperator*{\argmax}{argmax}
\newtheorem{definition}{Definition}
\newtheorem{theorem}{Theorem}

\newtheorem{conjecture}{Conjecture}

\newcommand{\Pro}{\mathbb{P}}
\crefname{appsec}{Appendix}{Appendices}
\newcommand{\crefapp}[1]{\ifdefined\LONG \cref{#1} \else the extended version of the paper \fi
}

%% file: main.tex
\title{Exact Recovery for a Family of Community-Detection Generative Models} 

\author{\IEEEauthorblockN{Luca Corinzia,
Paolo Penna, Luca Mondada and Joachim M. Buhmann}
\IEEEauthorblockA{
Department of Computer Science \\
ETH Zürich, Switzerland\\
\{luca.corinzia,paolo.penna,jbuhmann\}@inf.ethz.ch,  lmondada@student.ethz.ch}}

\maketitle

\begin{abstract}
\input{abstract.tex}
\end{abstract}

\ifdefined\LONG \else
\textit{A full version of this paper is accessible at:}
\url{https://arxiv.org/abs/1901.06799}

\fi
\section{introduction}
\subsection{Motivation and main contributions}
\input{motivation.tex}
\subsection{Related work}
\input{related_work.tex}
\section{Planted Random Energy Model}
\label{sec:p-rem}
\input{planted_REM.tex}
\section{Weighted Stochastic Block Model}
\label{sec:w-sbm}
\input{planted_k_subgraph.tex}

\section{The 2-WSBM on Hypergraphs}
\label{sec:w-hsbm}
\input{generalization.tex}
\section{Conclusion and future work}
\input{conclusion.tex}
\paragraph*{Acknowledgment}
We thank Wojciech Szpankowski for spotting an error in an early version this work, and Alexey Gronskiy and Andreas Krause for  valuable comments.

%% file: abstract.tex
Generative models for networks with communities have been studied extensively for being a fertile ground to establish information-theoretic and computational thresholds. In this paper we propose a new toy model for planted generative models called planted Random Energy Model (REM), inspired by Derrida's REM. For this model we provide the asymptotic behaviour of the probability of error for the maximum likelihood estimator and hence the exact recovery threshold. As an application, we further consider the 2 non-equally sized community Weighted Stochastic Block Model (2-WSBM) on $h$-uniform hypergraphs, that is equivalent to the P-REM on both sides of the spectrum, for high and low edge cardinality $h$. We provide upper and lower bounds for the exact recoverability for any $h$, mapping these problems to the aforementioned P-REM. To the best of our knowledge these are the first consistency results for the 2-WSBM on graphs and on hypergraphs with non-equally sized community.

%% file: motivation.tex
Random combinatorial optimization problems have been subject of intensive research in recent years in various disciplines, including statistical mechanics, combinatorial optimization and information theory \cite{mezard2009information}. 
A fruitful toy-model for random combinatorial optimization problem is the \emph{Random Energy Model (REM)} that assumes configuration with \emph{independent identically distributed} weights. Despite being a very simplistic model that does not show any spin glass behaviour, it has been used as a comparison to other random combinatorial optimization problems in the community-detection field, in those regimes where solutions are weakly correlated in the thermodynamical limit \cite{gronskiy2018free,buhmann2017phase,buhmann2014free}. 

In this line of research we define a new generative model called planted-REM (P-REM) inspired by the REM mentioned above, and embed it into a family of generative models for the community detection problem on hypergraphs. 
These models generate random instances of the form ``signal+noise'' as follows: first some randomly chosen solution is \emph{planted} and then Gaussian \emph{noise} is added to the instance. 
The P-REM model is the planted counterpart of REM where solutions have independent random weights, and one planted solution has a bias $\mu>0$. In the generative models for hypergraphs, the planted solution is a cluster or community of $k$ random nodes, and the weights of all hyperedges within this community have a bias $\mu>0$. Unlike the P-REM model, solutions which share some edges are statistically \emph{dependent}. A fundamental question for these models is whether the planted solution can be \emph{recovered} despite the noise.

The parameter $h$ that defines the edge cardinality in the hypergraph regulates how much different solutions are statistically correlated. On the two sides of the spectrum, i.e. $h=k$  and $h=1$, the model ``collapses'' to the P-REM with statistically \emph{independent} solutions, as each solution contains exactly one hyperedge. For intermediate values of $h$ we have the 2-Clusters Weighted Stochastic Block Model (2-WSBM) on hypergraphs with \emph{highly correlated} solutions: two solutions with large overlapping nodes share many edges and thus have many common random variables. The 2-WSBM and in general the SBM is a well studied model in social networks where the edge (and the edge weights) are generated at random in a way that reflects the membership of nodes to (unknown) communities. The main question for this model is whether it is possible to recover the community from the edge weights.

Interestingly, since in  the P-REM model ($h=k$) solutions are independent, finding the optimum requires 
searching through all $\binom{N}{k}$ solutions. On the contrary, for small $h\ll k$, the optimum might in principle be computed faster 
by exploiting the dependency between solutions. 
The contributions of this paper and its outline can be summarized as follow:
\begin{itemize}
\item In \Cref{sec:p-rem}, we introduce a generative model for planted REM, and establish necessary and sufficient conditions for the asymptotic recoverability of the planted solution.
\item In \Cref{sec:w-sbm} and \Cref{sec:w-hsbm}, we define the 2-WSBM on graphs and on generic $h$-uniform hypergraph and provide general recoverability condition for any $h$, summarized in \Cref{tab:summary}. 
\item We provide a technique to map the event of success recovery of these combinatorial problems to that of a corresponding P-REM. In this sense, this approach provides a \emph{general technique} which can be of relevance also in other applications. 
\end{itemize}

%% file: related_work.tex
Historically the first approaches to the random combinatorial optimization problems were driven by the interest on disordered system and spin glass behaviour in the statistical mechanics community. The REM \cite{derrida1981random,derrida1980random} was first introduced as a solvable mean field model of a disordered system.  This work inspired several follow ups, like the Sherrington-Kirkpatric model for spin glass, the statistical mechanics approach to the travel salesman problem and others
\cite{mezard1987spin}.

These problems also raise often as Maximum Likelihood (ML) estimators in generative models for graph. In these models ML estimators are used to define information-theoretic conditions under which the recovery of the generative model parameters can or cannot be solved irrespective of complexity or algorithmic considerations. In the following we give an overview on a few generative models for community detection on graphs that has been proposed so far.

In the \emph{planted clique problem}, an \emph{unweighted} graph is generated according to a \emph{semi-random} model where the edges are drawn according to Bernulli distributions, i.e., the standard Erd\"os-R\'enyi random graph (ER). In the simplest version, a set $S$ of nodes is chosen at random as the planted clique and all edges connecting them are added to the graph, while every other edge is included with probability $1/2$. This version exhibits a \emph{computational-statistical gap} \cite{steinhardt2017does}, meaning that despite it is possible to recover planted cliques of size $|S|\geq 2 \log_2 n$, the best known algorithms require $|S|\geq n^{1/2}$ \cite{barak2016nearly}. 

The SBM is another well known generative model for (multi) community detection. In its classic form, the SBM is another generalization of the ER, where each vertex belongs to one of $g$ groups, and each undirected edge is drawn according to a probability distribution that depends only on the group memberships of the relative vertices \cite{holland1983stochastic}. The classic SBM exhibits no statistical gap, and the recovery thresholds have been found recently for symmetric \cite{abbe2016exact,mossel2015consistency} and non-symmetric \cite{abbe2015community} model. Despite these achievements, open problems still exist for the various SBM generalizations, like weighted-SBM (WSBM) and SBM on (homogeneous) hypergraph (hSBM).
WSBM includes in the model additional information on the edges (represented by the weights) that can be used to better detect the communities \cite{aicher2013adapting,aicher2014learning,peixoto2018nonparametric}. The only information-theoretic result on WSBM is given in \cite{jog2015information} where the exact recovery threshold is given for a homogeneous WSBM with exactly equally sized communities.
The SBM on hypergraph has been instead proposed to model various grouping problems arising from computer vision and signal processing where only a multi-similarity is available as a function of more then just two points \cite{chien2018community,ghoshdastidar2014consistency}. The first consistency result for these models has been given in \cite{ghoshdastidar2014consistency} and on the follow-up \cite{ghoshdastidar2017consistency} for the special case of spectral algorithms respectively on uniform and non-uniform hypergraphs. Recently \cite{kim2017community} studied for the first time the weighted SBM on hypergraphs, providing recoverability thresholds for the homogeneous case with equally sized community.

%% file: planted_REM.tex
In this section we define the P-REM as a toy model for planted generative models on graphs. We assume here $k$ and $M$ to be respectively the number of biased (planted) and unbiased Gaussian random variables.
\begin{definition}\label{def:P-REM}
Let $M$ be a positive integer, $\mu$ and $\sigma$ two positive constants. The couple $(S,\mathbf{E})$, with $\mathbf{E} = (E_1, \dots, E_{M+k})$, is drawn under P-REM$(\mu,\sigma ,M,k)$ (denoted in the following also as $k$-P-REM) if the states $E_i$ are random variable \emph{conditional independent given $S$} and normally distributed as:
\begin{equation}
\begin{split}
E_{i} &\sim 
\begin{cases}
\mathcal{N}\left(\mu,\sigma^2\right) \quad i \in S \\
\mathcal{N}\left(0,\sigma^2\right) \quad i \notin S
\end{cases}
\end{split}
\end{equation}
where $S \subset \{1,\dots,M+k\}$, $|S| = k$ is drawn uniformly at random.
\end{definition}
The Gaussian weights are used for historical reasons, and a natural extension of this work can account for any arbitrary distribution, like in the work on WSBM in \cite{aicher2013adapting}. Note also that in the statistical mechanics literature the number of states is denoted as $M = 2^N$ to resemble a $1/2$-spin model.
\begin{definition}
Exact recovery for the P-REM is achieved if there exists an algorithm that takes $\mathbf{E}$ as input and outputs $\hat{S} = \hat{S}(\mathbf{E})$ such that
\[
\Pro[S = \hat{S}] = 1 - o_M(1)
\]
\end{definition}
To establish the information-theoretic limit for the P-REM we have to study the algorithm the maximizes the probability of recovery the correct biased index, that is the Maximum A Posteriori (MAP) decoding. In the case of the P-REM, the indices of biased weights are drawn uniformly, hence the MAP estimator corresponds to the ML estimator, that coincides to the indices of the top (largest) $k$ weights. 
\begin{theorem}\label{theo:rem_ML}
The ML estimator for the P-REM reads 
\[
\hat{S}^{ML}(\mathbf{E}) = \argmax_{\substack{\hat{S} \subset \{1,\dots,M+k\} \\ |\hat{S}| = k}} \sum_{j \in \hat{S}} E_j
\]
\end{theorem}
\begin{IEEEproof}
See \crefapp{app:rem}.
\end{IEEEproof}

Now we can establish the information-theoretic limit for exact recovery in the P-REM. We first study the recoverability condition for the $1$-P-REM, that depends intuitively on the magnitude of the signal to noise ratio (SNR) of the model. Then we will extend the results for any $k \le N^{\alpha}$ with $\alpha<1$. 

\begin{theorem}
\label{theo:rem_recovery}
Given a P-REM with parameters $(\hat{\mu}\log M, \hat{\sigma}\sqrt{\log M/2},M,1)$ the recovery probability of the ML estimator has asymptotics
\begin{equation}
\label{eq:REM_asymptothics}
\mathbb{P}\left[\hat{\alpha} = \alpha\right] = 
\begin{cases}
\begin{aligned}
&o(1) \quad &\gamma < 1 \\
&1 - \frac{ 1}{M^{(\gamma - 1)^2 + o(1)} }  \quad &1 < \gamma < 2 \\
&1 - \frac{1}{ M^{(\frac{\gamma^2}{2} - 1) + o(1) }} \quad &2 < \gamma
\end{aligned}
\end{cases}
\end{equation}
where $\gamma = \hat{\mu}/\hat{\sigma}$ is the SNR. Hence exact recovery is solvable if $\gamma > \gamma_c$ and unsolvable if  $\gamma < \gamma_c$, with $\gamma_c = 1$.
\end{theorem}

\begin{IEEEproof}
Without loss of generality let assume $l$ to be the index of the planted solution. The probability for the ML estimator to correctly identify the planted solution reads:
\begin{align*}
\mathbb{P}& \left[ \hat{l} = l\right] =\mathbb{P}\left[E_{l} = \max_{i=1,\dots,M +1} E_i\right] \\ 
&= \mathbb{P}\left[E_{l} \ge E_i, \ \ \forall \  i=1,\dots,M +1 \right] \\
&= \mathbb{E}_{E_{l}} \mathbb{P}\left[E_{l} \ge E_i, \ \ \forall \ i=1,\dots,M+1 | E_{l}\right] \\
&= \mathbb{E}_{E_{l}} \prod_{i=1,i \ne l}^{M+1} \mathbb{P}\left[E_{l} \ge E_i | E_{l} \right] =\\
&= \mathbb{E}_{E_{l}} \prod_{i=1,i \ne l}^{M+1} \phi\left(\frac{E_{l}\sqrt{2}}{\sqrt{\log M} \hat{\sigma}} \right) = \mathbb{E}_{E_{l}} \phi\left(\frac{E_{l}\sqrt{2}}{\sqrt{\log M} \hat{\sigma} } \right)^{M}
\end{align*}
where $\phi(x)$ is the cumulative distribution function (cdf) of the standard Gaussian distribution. Now we can study the thermodynamical limit $M \to \infty$.
Let us perform a change of variable $E_{l} = \epsilon \log M \hat{\sigma}$ and let us call $\gamma$ the SNR as $\gamma = \hat{\mu}/\hat{\sigma}$. Then the probability of success reads
\begin{equation}
\label{eq:rem_succ}
\sqrt{\frac{\log M}{\pi}} \int_{-\infty}^{+\infty} d\epsilon \  \phi\left(\sqrt{2\log M}\epsilon\right)^{M} e^{-\log M(\epsilon-\gamma)^2}
\end{equation}
Note that we can not apply the method of steepest descent \cite{wong2001asymptotic} for the integral in this form since the dependency on $M$ in not only on the exponential term but also on the function to be averaged. The idea of the proof is that the Gaussian probability distribution for large $M$ converges to a delta function centered in $\gamma$ and the function to be averaged, namely $\phi\left(\sqrt{2\log M}\epsilon\right)^{M}$, converges to a (Heaviside) step function in $1$ in the same regime. This gives intuitively the recovery threshold $\gamma_c = 1$. The corrections to $1$ in the high $\gamma$ regime showed in \cref{eq:REM_asymptothics} are given by (i) the Gaussian cdf for $1 < \gamma < 2$ and (ii) the correction from $1$ of the step function for $\gamma > 2$. See \crefapp{app:rem} for further details.

\textbf{Comment}: Note that the behaviour of $\phi\left(\sqrt{2\log M}\epsilon\right)^{M}$ is given by the \emph{Fisher–Tippett–Gnedenko} theorem (see \cite{welsch1973convergence} and \crefapp{app:rem}) in a particular regime, namely the case $\phi\left(a_M x + b_M\right)^{M} \approx exp(-e^{-x})$ (Gumbel cdf) for any given finite $x$ and $a_M \approx \frac{1}{(2 \log M)^{\frac{1}{2}}}$, $b_M \approx (2\log M)^{\frac{1}{2}}$. Hence it is easy to see that the function here considered $\phi\left(\sqrt{2\log M}\epsilon\right)^{M}$ behaves like a heaveside step function for $\epsilon >1$ and $\epsilon < 1$ and instead behaves like the Gumbel  function close to $1$, for $\epsilon \approx 1 + \frac{x}{2 \log M}$. More details in the \crefapp{app:rem}.
\end{IEEEproof}

For the general P-REM the same recoverability threshold of the $1$-P-REM applies as long as $k$ grows slower then any power of $M$. For $k = M^{\alpha}$ and $\alpha$ a constant strictly positive ($0<\alpha < 1$) we have a gap of size $\sqrt{\alpha}$, as shown in the following theorem:
\begin{theorem}
\label{theo:k_p_rem_recovery}
A P-REM problem with parameters $(\hat{\mu}\log M, \hat{\sigma}\sqrt{\log M/2},M,k)$ is solvable for any constant $\gamma>1 + \sqrt{\alpha}$ and unsolvable for $\gamma<1$, where $\alpha$ is the smallest value such that $k \leq M^\alpha$ and $0 < \alpha < 1$.
\end{theorem}
\begin{IEEEproof}
See \crefapp{app:rem}.
\end{IEEEproof}

%% file: planted_k_subgraph.tex
In this section we define the weighted Stochastic Block Model (WSBM) with \emph{two clusters} and gaussian weights. In particular let us consider a \emph{complete} graph $\mathcal{G}$ with set of nodes $\mathcal{V} = \{1,\dots,N\}$.
\begin{definition}
\label{def:wsbm}
Let $N$ and $k$ two positive integers, $\mu$ and $\sigma$ two positive constants. The couple $(S,E)$ is drawn under 2-WSBM$(\mu,\sigma ,N,k)$ if the edge weights $E_{ij}$ are random variables \emph{conditional independent given $S$} and normally distributed as:
\begin{equation*}
\begin{split}
E_{ij} \sim 
\begin{cases}
\mathcal{N}\left(\mu,\sigma^2\right) \quad \ &i,j \in S \\
\mathcal{N}\left(0,\sigma^2\right) \quad &otherwise
\end{cases}
\end{split}
\end{equation*}
where $S \subset \mathcal{V}$, $|S| = k$, is drawn uniformly at random.
\end{definition}

Note that this definition is equivalent to the definition of a \emph{non-homogeneus} 2-WSBM given in \cite{aicher2013adapting}, in the case of Gaussian weights. Formally the cluster connectivity matrix of the model described in \Cref{def:wsbm} according to the framework given in \cite{aicher2013adapting} reads 
\begin{equation*}
W =
\begin{bmatrix}
p & q \\
q & q \\
\end{bmatrix}
\end{equation*}
where $p = \mathcal{N}\left(\mu,\sigma^2\right)$ is the pdf relative to the weights in the planted cluster $S$ and $q = \mathcal{N}\left(0,\sigma^2\right)$ is the pdf relative to both the weights of the cluster $\mathcal{V} \setminus S$ and the weights across the two clusters. Note also that in the definition here used we consider the size of one cluster to be $|S| = k$, hence the size of the other cluster is $|\mathcal{V} \setminus S| = N-k$.
\begin{definition}\label{def:recovery:subgraph}
Exact recovery for the 2-WSBM in \Cref{def:wsbm} is achieved if there exists an algorithm that takes $E$ as input and outputs $\hat{S} = \hat{S}(E)$ such that
\[
\Pro[S = \hat{S}] = 1 - o_N(1)
\]
\end{definition}

As for the case of the P-REM, the information-theoretic recoverabilty limit is given by the MAP estimator, that for clique nodes $S$ drawn at random corresponds to the ML estimator, that is described in the following theorem.  

\begin{theorem}
\label{theo:subgraph_ML}
The ML estimator for the 2-WSBM in \Cref{def:wsbm} is the densest $k$-subgraph.
\begin{equation*}
\hat{S}^{ML}(E) =  \argmax_{\hat{S} \colon |\hat{S}| = k} W(\hat{S})
\end{equation*}
where $W(\cdot)$ is the solution weight and is defined as $W(A) := \sum_{i,j \in A} E_{ij}$
\end{theorem}

\begin{IEEEproof}
See \crefapp{app:k_subgraph}.
\end{IEEEproof}

Also in the 2-WSBM the information-theoretic limit for exact recovery on the SNR, and the exact thresholds are given by the following theorem.

\begin{theorem}
\label{theo:k_subgraph_recovery}
Exact recovery in the 2-WSBM with parameters $(\hat{\mu}\log N, \hat{\sigma}\sqrt{\log N/2},N,k)$ is unsolvable if $\gamma< \gamma_{-} = \sqrt{\frac{1}{k-1}}$ and solvable if  $\gamma > \gamma_{+}$, where the upper threshold is defined according to the different $k$ regimes as:
\begin{align*}
\gamma_{+} = \begin{cases}
2 \sqrt{\frac{1}{k-1}} \quad & k = o(\log N) \\
2 \sqrt{\frac{1 + \log 2 + \frac{1}{c}}{\log N}} \quad & \frac{k}{\log N} \to c, c \in \mathbb{R}^+ \cup \{+\infty\} \\
2 \sqrt{\frac{1 + \log 2}{(1-\alpha)\log N}} \quad & k \lessapprox N^{\alpha}, 0<\alpha<1
\end{cases}
\end{align*}
\end{theorem}

\begin{IEEEproof}[Proof (Main Idea)]
\textbf{Lower Bound}: We shall reduce the model to the 1-P-REM problem by partitioning the set of all solutions into groups, where the solutions in each group form an instance of 1-P-REM. We then use the bounds for 1-P-REM to show that the probability that the planted solution ``fails'' against the solutions in a single group is sufficiently small. By taking the union bound over all groups, we get an upper bound on the fail probability that is vanishing (hence a lower bound on the recovery probability that converges to 1). 

The actual partition of the solutions is done in order to guarantee that, within each group, solutions are  ``statistically independent'' as in the P-REM (see \Cref{fig:main-idea:theo:k_subgraph_recovery}).
\begin{figure}
\centering
\includegraphics[scale=.45]{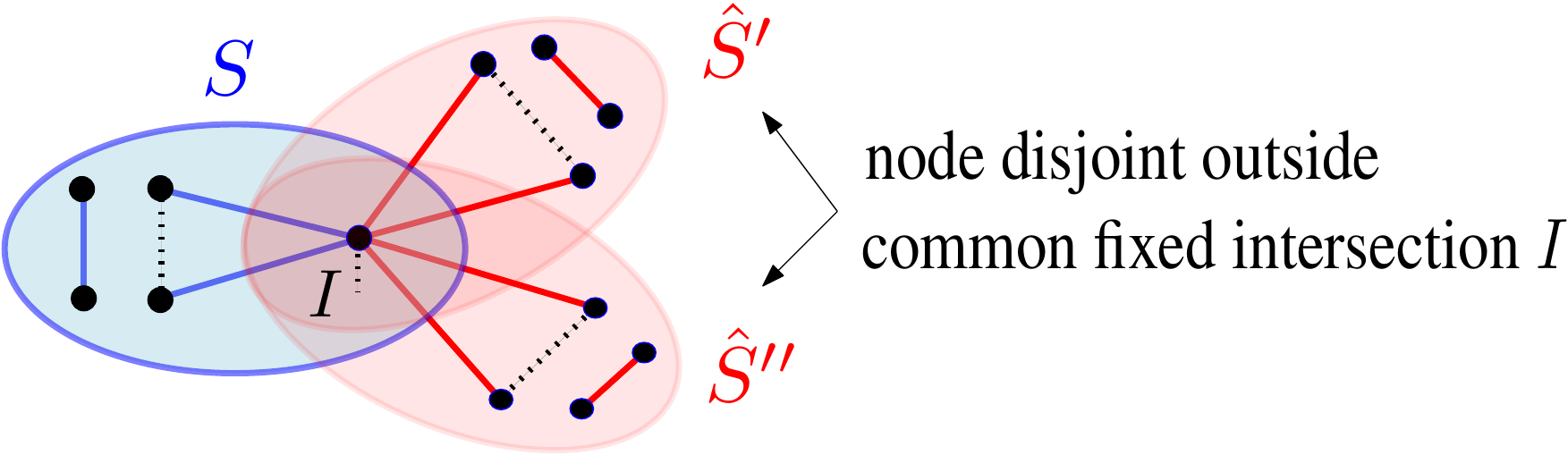}
\caption{The main idea of the proof of \Cref{theo:k_subgraph_recovery} is the reduction to the P-REM problem: fix an intersection $I$ and build a P-REM with ''node-disjoint'' solutions.}
\label{fig:main-idea:theo:k_subgraph_recovery}
\end{figure}
Given the planted solution $S$, fix a possible intersection, i.e. a subset $I$ of $m$ nodes in $S$ and consider all solutions $\hat S$ consisting of nodes in this intersection $I$ and of $k - m$ nodes not in $S$. Then, further partition these solutions into groups such that,  each group $\mathcal C_{\hat S}$ consists of $M_m = \lfloor \frac{N- k}{k - m}\rfloor$ solutions that are node-disjoint except the common intersection $I$ (simply take disjoint blocks of $k-m$ nodes each). The key point is that the nodes in $I$ are contained in all solutions and in the planted solution as well. We can therefore remove all these edges from the consideration, and restrict to the remaining edges with weights $W_{REM}(\hat S) := W(\hat{S}) - W(I)$ for any solution $\hat S$ as above (including the planted solution $S$). 
Note that the following holds: (i) Each $W_{REM}(\hat S)$ is a Gaussian random variable which is given by the sum of $\ell_m = \binom{k}{2} - \binom{m}{2}$ edge weights. (ii) Since two solutions $\hat S'$ and $\hat S''$ do not share any edge other than those inside $I$, $W_{REM}(\hat S')$ and $W_{REM}(\hat S'')$ are independent Gaussian random variables. (iii) $W(S) > W(\hat S)$ if and only if $W_{REM}(S) > W_{REM}(\hat S)$.
Hence variables in $\mathcal C_{\hat S}$ are a 1-P-REM with $1+M_m = 1+ |\mathcal C_{\hat S}|$ states where each state (solution) is a Gaussian random variable distributed as  
\begin{equation*}
\begin{split}
E_{\hat{S}}=W_{REM}(\hat{S}) \sim \begin{cases}
\mathcal{N}\left(\ell_m \mu,\ell_m \sigma^2\right) \quad  &\hat{S} = S \\
\mathcal{N}\left(0,\ell_m\sigma^2\right) \quad &\hat{S} \neq S
\end{cases}
\end{split}
\end{equation*}
Our construction yields $M_m = \lfloor\frac{N-k}{k -m}\rfloor \ge N^{1-\alpha}$ and the resulting 1-P-REM has $\mu_{REM}=\ell_m \mu$ and $\sigma^2_{REM}=\ell_m \sigma^2$, which results in a SNR  $\gamma_m = \frac{\hat \mu_{REM}}{\hat \sigma_{REM}}   \approx \frac{\hat \mu}{\hat \sigma} \sqrt{\ell_m } = \gamma \sqrt{\ell_m }$. 
For $\gamma> \gamma_{+}$, it follows that $\gamma_m>2$ for all $m$,  and \cref{eq:REM_asymptothics} yields a probability of ``failing'' against some solution in $\mathcal{C}_{\hat S}$ which is ``sufficiently small'' to apply the union bound over all possible $m$ and all possible $\mathcal{C}_{\hat S}$ necessary to cover all solutions.

\textbf{Upper Bound}: We construct a single $\mathcal{C}_{\hat{S}}$ with maximal $m=k-1$ and observe that, since $\ell_{k-1} = k-1$, for $\gamma< \gamma_{-} = \sqrt{\frac{1}{k-1}}$  the signal to noise ratio of the corresponding P-REM satisfies $\gamma_{k-1}<1$. By \cref{eq:REM_asymptothics} the probability of ``failing'' against some solutions in $\mathcal{C}_{\hat S}$ tends to $1$, and the probability of failing against an arbitrary solution is lower bounded by it. See \crefapp{app:k_subgraph} for the full proof.
\end{IEEEproof}

Interestingly the exact recovery results given in \cite{jog2015information} for the \emph{homogeneous} 2-WSBM with \emph{exactly equally sized community} is based on the $\frac{1}{2}$-Rényi divergence between the pdfs of the inter-cluster and intra-cluster weights, that for the gaussian case is an explicit function of the SNR. Hence, despite the two models considered in this paper and in \cite{jog2015information} are distinct, the same parameter regulates their recoverability conditions. Notice also that our scenario is \emph{complementary} to that given in \cite{jog2015information} since the equally sized community constraint is equivalent to $k = N$, hence $\alpha = 1$.

From \Cref{theo:k_subgraph_recovery} we can see that the recoverability threshold decreases with $k$ as $\sqrt{\frac{1}{k-1}}$ up to $k$ of the magnitude of $\log N$, so that it is easier (we can afford lower SNR) to detect larger communities. This result is intuitive and it is in line with what has been found already for the planted-clique problem \cite{steinhardt2017does}, where planted cliques can be recovered only on the regime $|S| = k \ge 2 \log_2 N$. For larger $k$ the threshold decreases with $k$ more slowly, only through the parameter $\alpha = \frac{\log k}{N}$.

\begin{conjecture}
In the simple case in which $k = 2$ ($\alpha = 0$), the 2-WSBM is exactly equivalent to the 1-P-REM with number of states $M = \binom{N}{2} \approx N^2$, hence it has a critical $\gamma_c = \sqrt{2}$. This justifies the conjecture, left for future development, that the critical gamma for this 2-WSBM with $k = o(\log N)$ is $\gamma_c = \sqrt{\frac{2}{k-1}}$.
\end{conjecture}

%% file: generalization.tex
In this section, we consider the generalization of the 2-WSBM to the complete $h$-uniform hypergraph. We consider hyperedges of uniform edge-cardinality $h$. Formally:

\begin{definition}[2-hWSBM]
\label{def:hWSBM}
Let $N$, $k$, and $h$ be positive integers, $\mu$ and $\sigma$ two positive constants. The couple $(S,E)$ is drawn under 2-hWSBM$(\mu,\sigma ,N,k,h)$ if the weights  $E_{i_1 i_2 \cdots i_h}$ are random variables \emph{conditional independent given $S$} and normally distributed as:
\begin{equation*}
\begin{split}
E_{i_1 i_2 \cdots i_h} \sim 
\begin{cases}
\mathcal{N}\left(\mu,\sigma^2\right) \quad &\ i_1, i_2, \ldots, i_h \in S \\
\mathcal{N}\left(0,\sigma^2\right) \quad &otherwise
\end{cases}
\end{split}
\end{equation*}
where $S \subset \mathcal{V}$, $|S| = k$, is drawn uniformly at random.
\end{definition}
	
The corresponding exact recovery problem is defined as for the 2-WSBM (\Cref{def:recovery:subgraph}). Also, \Cref{theo:subgraph_ML} extends easily to this model, hence the ML estimator yields the densest $k$-sub-hypergraph. The recoverability condition for generic $h$ are given by the following:
\begin{theorem}
\label{theo:k_subhypergraph_recovery}
The 2-hWSBM, with $2 \leq h \leq k$ and with parameters $(\hat{\mu}\log N, \hat{\sigma}\sqrt{\log N/2},N,k,h)$, is unsolvable if $\gamma< \gamma_{-} = \sqrt{\frac{1}{\binom{k-1}{h-1}}}$ and solvable if $\gamma > \gamma_{+}$, where the upper threshold is defined according to the $k$ and $h$ regimes as:
\begin{align*}
\gamma_{+} = \begin{cases}
2 \sqrt{\frac{\frac{h}{2}}{\binom{k-1}{h-1}}} \ & \frac{1}{h} \binom{k-1}{h-1} = o(\log N) \\
2 \sqrt{\frac{1 + \log 2 + \frac{1}{c}}{\log N}} \ & \frac{1}{h} \binom{k-1}{h-1} / \log N \to c, c \in \mathbb{R}^+ \cup \{+\infty\} \\
2 \sqrt{\frac{1 + \log 2}{(1-\alpha)\log N}} \ & k \lessapprox N^{\alpha}, 0<\alpha<1
\end{cases}
\end{align*}
\end{theorem}
	
\begin{IEEEproof}
See \crefapp{app:hypergraph}.
\end{IEEEproof}
	
Note that for $h=2$, this yields exactly the results in \Cref{theo:k_subgraph_recovery}, and also that the conjecture extends easily to the case $k=h$, for which we recover the P-REM with number of states $M = \binom{N}{k} \approx N^k$. The model has thereof an exact threshold $\gamma_c = \sqrt{k}$, hence we conjecture that the threshold for the general problem on $h$-hypergraphs reads $\gamma_c = \sqrt{\frac{h}{\binom{k-1}{h-1}}}$. The other side of the $h$ range is defined by $h=1$ (not included in \Cref{theo:k_subhypergraph_recovery}), for which it is easy to show that the corresponding 2-WSBM is equivalent to a $k$-PREM with number of states $M=N$. Hence the thresholds are given by \Cref{theo:k_p_rem_recovery}. A summary of all the results of the paper for the regime $k = o(\log N)$ is given in \Cref{tab:summary}.
Note that also in this case previous results found in \cite{kim2017community} on 2-hWSBM are orthogonal to ours since those are given for \emph{equally sized community} and \emph{homogeneous} models.
\begin{table}[htbp]
\renewcommand{\arraystretch}{1.3}
\caption{Thresholds for the 2-$h$WSBM at different $h$, and $k = o(\log N)$}
\label{tab:summary}
\centering
\begin{tabular}{|c||c|c|c|}
\hline
$h$ & Model & $\gamma_{-}$ & $\gamma_{+}$\\
\hline
\hline
$1$ & $k$-P-REM & $1$ & $1$ \\
$2$ & 2-WSBM & $\sqrt{\frac{1}{k-1}}$  & $2\sqrt{\frac{1}{k-1}}$ \\
$2 < h < k$ & 2-hWSBM & $\sqrt{\frac{1}{\binom{k-1}{h-1}}}$ & $2\sqrt{\frac{h/2}{\binom{k-1}{h-1}}}$ \\
$k$ & $1$-P-REM & 1 & 1 \\
\hline
\end{tabular}
\end{table}

%% file: conclusion.tex
Information-theoretic fundamental limits have been a fruitful approach to establish benchmarks for algorithm performance. A important example is given by Shannon’s coding theorem that gives a recoverability threshold for coding algorithms, located at the channel capacity, and that gave rise to a plethora of coding algorithms in the last decades. Recently the same approach has been carried out in the area of clustering and community detection in (hyper)graphs. Motivated by this we established a new toy model for generative model called (k)-planted REM giving a description of the maximum likelihood failing probability in the asymptotic limit and consequently a recoverability threshold. We embedded this model in the well know (weighted) Stochastic Block model framework for community detection in h-uniform hypergraphs. These models correspond to the planted REM for both sides of the $h$ spectrum, $h = k$ and $h = 1$. For all these models we provided the first recoverability conditions. As future research directions, we plan to provide matching thresholds for the 2-WSBM on $h$-uniform hypergraphs for any given $h$ and to account for natural extensions, like multi-community and non Gaussian probability distributions.

%% file: planted_rem_appendix.tex
\begin{IEEEproof}[Proof of \cref{theo:rem_ML}]
\begin{align*}
\Pro_e &= \Pro_{S,\mathbf{E}}\left[\hat{S}(\mathbf{E}) \ne S\right] = \\
& =\sum_{S} \int d\mathbf{E} \ p(S) p(\mathbf{E} | S) \mathbbm{1}[\hat{S}(\mathbf{E}) \ne S]
\end{align*}
hence the algorithm the minimizes the error is the one the maximizes $p(\mathbf{E} | S) \mathbbm{1}[\hat{S}(\mathbf{E}) = S]$ for every set $S$ and every configuration $\mathbf{E}$. 
\begin{align*}
\hat{\alpha}^{MAP}(\mathbf{E}) &= \hat{\alpha}^{ML}(\mathbf{E}) \\
& = \argmax_{\substack{\hat{S} \subset \{1,\dots,M+k\} \\ |\hat{S}| = k}} p(\mathbf{E} | \hat{S}) \\
& = \argmax_{\substack{\hat{S} \subset \{1,\dots,M+k\} \\ |\hat{S}| = k}} \prod_{i \in \hat{S}} \mathcal{N}\left(E_i|\mu,\sigma^2\right)
\prod_{j \notin \hat{S}} \mathcal{N}\left(E_j| 0,\sigma^2\right)  \\
&= \argmax_{\substack{\hat{S} \subset \{1,\dots,M+k\} \\ |\hat{S}| = k}} \prod_{i \in \hat{S}} \frac{\mathcal{N}\left(E_i| \mu,\sigma^2\right)}{\mathcal{N}\left(E_i| 0,\sigma^2\right)} \prod_{j}
\mathcal{N}\left(E_j| 0,\sigma^2\right) \\
&= \argmax_{\substack{\hat{S} \subset \{1,\dots,M+k\} \\ |\hat{S}| = k}} \sum_{i \in \hat{S}} E_i
\end{align*}
\end{IEEEproof}

\begin{IEEEproof}[Proof of \cref{theo:rem_recovery}]
Let us first observe the behavior of $\phi(x)$ for large x:
\begin{equation*}
\phi(x) =
\begin{cases}
\begin{aligned}
1 - \frac{1}{\sqrt{2\pi}x} e^{-\frac{x^2}{2}} (1+ \mathcal{O}(1/x^2))  \quad &for \ x \to +\infty \\
\frac{1}{\sqrt{2\pi}|x|}e^{-\frac{x^2}{2}} (1+ \mathcal{O}(1/x^2)) \quad &for \ x \to -\infty
\end{aligned}
\end{cases}
\end{equation*} 
hence as a function of $\epsilon$ we have:
\begin{equation*}
\begin{split}
\phi&\left(\sqrt{2\log M} \epsilon\right) = \\
&= \begin{cases}
\begin{aligned}
& 1-\frac{e^{-\log M\epsilon^2}}{2\sqrt{\pi \log M}\epsilon} \left(1+ \mathcal{O}\left(\frac{1}{\log M \epsilon^2}\right)\right) \ & \forall \ &\epsilon >0 \\
& \frac{e^{-\log M\epsilon^2}}{2\sqrt{\pi \log M}|\epsilon|} \left(1+ \mathcal{O}\left(\frac{1}{\log M \epsilon^2}\right)\right) \ & \forall \ &\epsilon <0
\end{aligned}
\end{cases}
\end{split}
\end{equation*} 
where the asymptotic notation is referred to $M \to +\infty$, here and in the rest of this proof.
For $\epsilon > 0 $ then we have:
\begin{align}
\phi(&\sqrt{2\log M} \epsilon)^M = \exp(M \log (\phi(\sqrt{2\log M} \epsilon))) \notag \\
&= \exp\left(-M \frac{e^{-\log M\epsilon^2}}{2\sqrt{\pi \log M}\epsilon} \left(1+ \mathcal{O}\left(\frac{1}{\log M \epsilon^2}\right)\right) \right) \notag \\
&= \exp\left(- \frac{e^{\log M (1 - \epsilon^2)}}{2\sqrt{\pi \log M}\epsilon}  \left(1+ \mathcal{O}\left(\frac{1}{\log M \epsilon^2}\right)\right) \right) \label{eq:rem_phi_power_M}
\end{align}
then 
\begin{align}
\label{eq:rem_phi_behaviour}
\phi&(\sqrt{2\log M} \epsilon)^M = \notag \\
&=1 - \frac{e^{-\log M (\epsilon^2 - 1)}}{2\sqrt{\pi \log M}\epsilon} \left(1+ \mathcal{O}\left(\frac{1}{\log M (\epsilon^2-1)}\right)\right) \quad \epsilon > 1
\end{align}
For \cref{eq:rem_phi_behaviour} to hold we need $\log M (\epsilon^2 -1) \to +\infty$ hence it holds for any $\epsilon \ge 1 + \tau_M$ for which $\log M \tau_M \to +\infty$. Notice also that the asymptotics given in \cref{eq:rem_phi_behaviour} is equivalent to that given by the Fisher–Tippett–Gnedenko theorem for gaussian random variables \cite{welsch1973convergence} in the regime $\epsilon = 1 + \tau_M$ with $\tau_M = o(1)$ and $\log M \tau_M \to +\infty$.

We can now write the probability of success splitting the integral in \cref{eq:rem_succ} into the three intervals $(-\infty,0)$,$(0,1+\tau_M)$ and $(1+\tau_M,+\infty)$ where $\tau_M$ is defined as above.
\begin{align}
&\mathbb{P}\left[\hat{l} = l\right] = \notag \\
&= \sqrt{\frac{\log M}{\pi}} \int_{-\infty}^{0} d\epsilon \  \phi\left(\sqrt{2\log M}\epsilon\right)^{M} e^{-\log M(\epsilon-\gamma)^2} + \label{eq:rem_first_term}\\
&+ \sqrt{\frac{\log M}{\pi}} \int_{0}^{1+\tau_M} d\epsilon \  \phi\left(\sqrt{2\log M}\epsilon\right)^{M} e^{-\log M(\epsilon-\gamma)^2} + \label{eq:rem_second_term}\\
&+ \sqrt{\frac{\log M}{\pi}} \int_{1+\tau_M}^{+\infty} d\epsilon \  \phi\left(\sqrt{2\log M}\epsilon\right)^{M} e^{-\log M(\epsilon-\gamma)^2} \label{eq:rem_third_term}
\end{align}
Let us first consider the third term  (\cref{eq:rem_third_term}) that using \cref{eq:rem_phi_behaviour} reads:
\begin{align}
&
\!\begin{multlined}[t]
\sqrt{\frac{\log M}{\pi}} \int_{1+\tau_M}^{+\infty} d\epsilon \  \phi\left(\sqrt{2\log M}\epsilon\right)^{M} e^{-\log M(\epsilon-\gamma)^2} = \\
= \sqrt{\frac{\log M}{\pi}} \int_{1+\tau_M}^{\infty} d\epsilon \left( 1 - \frac{e^{-\log M (\epsilon^2-1)}}{2\sqrt{\pi \log M}\epsilon} (1+ \mathcal{O}(\frac{1}{\log M})) \right) \times \\
\times e^{-\log M(\epsilon-\gamma)^2} \end{multlined} \notag \\
&= \!\begin{multlined}[t]
\sqrt{\frac{\log M}{\pi}} \int_{1+\tau_M}^{\infty} d\epsilon \ e^{-\log M(\epsilon-\gamma)^2} + \\
- \sqrt{\frac{\log M}{\pi}} \int_{1+\tau_M}^{\infty} d\epsilon \ \frac{e^{-\log M (\epsilon^2-1)}}{2\sqrt{\pi \log M}\epsilon} e^{-\log M(\epsilon-\gamma)^2} \times \\
\times (1+ \mathcal{O}(\frac{1}{\log M}))
\end{multlined} \notag \\
&=
\left(1-\phi(\sqrt{2\log M} (1 + \tau_M-\gamma)) \right) +  \label{eq:rem_third_first} \\
&  \!\begin{multlined}[t] 
- \sqrt{\frac{\log M}{\pi}} \int_{1+\tau_M}^{\infty} d\epsilon \frac{e^{-\log M (\epsilon^2-1)}}{2\sqrt{\pi \log M}\epsilon} e^{-\log M(\epsilon-\gamma)^2}  \times \\
\times (1+ \mathcal{O}(\frac{1}{\log M})) \label{eq:rem_third_second}
\end{multlined}
\end{align}
The term in \cref{eq:rem_third_first} has been studied before, and approaches $0$ for $\gamma < 1$ and instead for $\gamma > 1$ reads: 
\begin{align*}
1- \phi(&\sqrt{2\log M}(1 + \tau_M - \gamma)) = \\ &=  1 - \frac{ e^{-\log M((\gamma - 1)^2 + o(1))}}{2 \sqrt{\pi \log M}(\gamma - 1) } (1+ \mathcal{O}(\frac{1}{\log M})).
\end{align*}
Now let us rewrite the term in \cref{eq:rem_third_second} as
\begin{align*}
&\sqrt{\frac{\log M}{\pi}} \int_{1+\tau_m}^{\infty} d\epsilon \frac{e^{-\log M (\epsilon^2-1)}}{2\sqrt{\pi \log M}\epsilon} e^{-\log M(\epsilon-\gamma)^2} = \\ &= \sqrt{\frac{\log M}{\pi}} \frac{e^{\log M}}{2\sqrt{\pi \log M}} \int_{1+\tau_M}^{\infty} d\epsilon \ \frac{1}{\epsilon} e^{-\log M(\epsilon^2 +(\epsilon-\gamma)^2 )} \\
&= \frac{e^{\log M (1 - \frac{\gamma^2}{2})}}{4\sqrt{\pi \log M}} 2 \sqrt{\frac{\log M}{\pi}} \int_{1+\tau_m}^{\infty} d\epsilon \ \frac{1}{\epsilon} e^{-2\log M(\epsilon -\frac{\gamma}{2})^2}
\end{align*}
Now we can see that for $\gamma > 2$ we can use steepest descent method \cite{wong2001asymptotic} to get:
\begin{multline*}
\frac{e^{\log M(1 - \frac{\gamma^2}{2})}}{4\sqrt{\pi \log M}} 2 \sqrt{\frac{\log M}{\pi}} \int_{1+\tau_M}^{\infty} d\epsilon \ \frac{1}{\epsilon} e^{-2\log M(\epsilon -\frac{\gamma}{2})^2} = \\ 
= \frac{e^{\log M (1 - \frac{\gamma^2}{2})}}{4\sqrt{\pi \log M}} \left(\frac{2}{\gamma} + \mathcal{O}(\frac{1}{\log M}) \right)
\end{multline*}
For  $1 < \gamma < 2$ we can see instead that 
\begin{equation*}
\begin{split}
&\frac{e^{\log M (1 - \frac{\gamma^2}{2})}}{4\sqrt{\pi \log M}} 2 \sqrt{\frac{\log M}{\pi}} \int_{1+\tau_M}^{\infty} d\epsilon \ \frac{1}{\epsilon} e^{-2\log M(\epsilon -\frac{\gamma}{2})^2}  \\
&\le \frac{e^{\log M(1 - \frac{\gamma^2}{2})}}{4\sqrt{\pi \log M}}  2 \sqrt{\frac{\log M}{\pi}} \int_{1+\tau_M}^{\infty} d\epsilon \ e^{-2\log M(\epsilon - \frac{\gamma}{2})^2} \\
&= \frac{e^{\log M (1 - \frac{\gamma^2}{2})}}{4\sqrt{\pi \log M}} \left(1- \phi\left(\sqrt{2\log M}\left(1 + \tau_M- \gamma/2\right) \right) \right) \\
&= \frac{e^{\log M (1 - \frac{\gamma^2}{2})} e^{-2\log M(\left(\frac{\gamma}{2} - 1\right)^2 + o(1) )}}{8 \pi \log M\sqrt{1 - \frac{\gamma}{2}}} (1+ \mathcal{O}(\frac{1}{\log M})) \\ 
&= \frac{e^{-\log M((\gamma + 1)^2+o(1))}}{8 \pi \log M \sqrt{1 - \frac{\gamma}{2}}} (1+ \mathcal{O}(\frac{1}{\log M}))
\end{split}
\end{equation*}
Hence the term in \cref{eq:rem_third_term} asymptotically reads: \begin{align*}
\sqrt{\frac{\log M}{\pi}} \int_{1}^{+\infty} d\epsilon \  \phi\left(\sqrt{2\log M}\epsilon\right)^{M} e^{-\log M(\epsilon-\gamma)^2} = \\ =
\begin{cases}
\begin{aligned}
&o(1) \quad &\gamma < 1 \\
&1 - \frac{ 1}{M^{(\gamma - 1)^2 + o(1)} }  \quad &1 < \gamma < 2 \\
&1 - \frac{1}{ M^{(\frac{\gamma^2}{2} - 1) + o(1) }} \quad &2 < \gamma
\end{aligned}
\end{cases}
\end{align*}
Now we show that the other two terms \cref{eq:rem_first_term} and \cref{eq:rem_second_term} do not change the asymptotic behaviour of the recovery probability. To see this note that $\phi\left(\sqrt{2\log M}\epsilon\right)^{M}$ is an increasing function of $\epsilon$ hence we can bound easily the two terms. The first term \cref{eq:rem_first_term} reads:
\begin{equation*}
\begin{split}
\sqrt{\frac{\log M}{\pi}} &\int_{-\infty}^{0} d\epsilon \  \phi\left(\sqrt{2\log M}\epsilon\right)^{M} e^{-\log M(\epsilon-\gamma)^2} \le \\
&\le \phi(0)^M \phi(-\sqrt{2\log M}\gamma ) = \frac{1}{2^M} \phi(-\sqrt{2\log M}\gamma ) \\ &=o\left(\frac{1}{M^\alpha}\right)
\end{split}
\end{equation*} 
for any constant $\alpha >0$, hence it is smaller than any contribution of term \cref{eq:rem_third_term} for $\gamma >1$. For the second term, let us sum up \cref{eq:rem_second_term} and \cref{eq:rem_third_first} to get the upper bound
\begin{equation*}
\begin{split}
&\!\begin{multlined}[t] 
\sqrt{\frac{\log M}{\pi}} \int_{0}^{1+\tau_M} d\epsilon \  \phi\left(\sqrt{2\log M}\epsilon\right)^{M} e^{-\log M(\epsilon-\gamma)^2} 
+ \\
+ 1 - \phi(\sqrt{2\log M}(1-\gamma ) )
\end{multlined}
\\
&\!\begin{multlined}[t]  \le 1 - (1- \phi(\sqrt{2\log M}(1+\tau_M))^M )(\phi(\sqrt{2\log M}(1-\gamma ) ) + \\
- \phi(-\sqrt{2\log M}\gamma ) \end{multlined} \\
&= 1 - \left(\frac{e^{-2\log M \tau_M}}{2\sqrt{\pi \log M}} + \mathcal{O}\left(\frac{1}{\log M}\right)  \right) \frac{1}{M^{(\gamma -1)^2 + o(1)}} \\
&= 1 - \frac{1}{M^{(\gamma -1)^2 + o(1)}}
\end{split}
\end{equation*} 
were the second last equality is given by \cref{eq:rem_phi_power_M} and the last equality is given by $\tau_M = o(1)$. Analogously we can get the lower bound 
\begin{equation*}
\begin{split}
&\!\begin{multlined}[t] 
\sqrt{\frac{\log M}{\pi}} \int_{0}^{1+\tau_m} d\epsilon \  \phi\left(\sqrt{2\log M}\epsilon\right)^{M} e^{-\log M(\epsilon-\gamma)^2} 
+ \\
+ 1 - \phi(\sqrt{2\log M}(1-\gamma ) 
\end{multlined}
\\
&\!\begin{multlined}[t]  \ge 1 - (1- \phi(0)^M )(\phi(\sqrt{2\log M}(1-\gamma ) ) + \\
- \phi(-\sqrt{2\log M}\gamma ) \end{multlined} \\
&= 1 - \frac{1}{M^{(\gamma -1)^2 + o(1)}}
\end{split}
\end{equation*} 
Summing up all the terms together we get:
\begin{equation*}
\mathbb{P}\left[\hat{l} = l\right] = 
\begin{cases}
\begin{aligned}
&o(1) \quad &\gamma < 1 \\
&1 - \frac{ 1}{M^{(\gamma - 1)^2 + o(1)} }  \quad &1 < \gamma < 2 \\
&1 - \frac{1}{ M^{(\frac{\gamma^2}{2} - 1) + o(1) }} \quad &2 < \gamma
\end{aligned}
\end{cases}
\end{equation*}
hence \cref{theo:rem_recovery} holds.
\end{IEEEproof}

\begin{IEEEproof}[Proof of \cref{theo:k_p_rem_recovery}]
\textbf{Lower bound}
Observe that exact recovery fails if and only if one biased weight is worst (smaller) than one of the $M+k$ unbiased weights. Then,  
\begin{align*}
\Pro[S \neq \hat S] &= \Pro\left[\bigcup_{i \in S} \{ E_{i} <  \max_{j \notin S}  E_{j} \} \right] 
\\
& \le k\Pro\left [E_{i} <  \max_{j \notin S} E_{j}\right]
\end{align*}
where the inequality is simply the union bound and $i \in S$. Now observe that the latter probability is just the probability of failing in a P-REM with $M$ states and thus the bounds in \eqref{eq:REM_asymptothics} for $1 < \gamma<2$ yield 
\[k\Pro\left [E_i <  \max_{j \notin \hat{S}} E_{j}\right] \leq \frac{k}{M^{(\gamma-1)^2}} \leq \frac{M^\alpha}{M^{(\gamma-1)^2}} =  o(1)
\]
since $\gamma>1 +\sqrt{\alpha}$ implies $(\gamma-1)^2>\alpha$. For $\gamma>2$, the corresponding bound in \eqref{eq:REM_asymptothics}  yields $k\Pro\left [E_i <  \max_{j \notin \hat{S}} E_{j}\right] \leq \frac{1}{M^{\frac{\gamma^2}{2}-2}}=o(1)$.

\textbf{Upper bound} For $\gamma<1$, the bounds in \eqref{eq:REM_asymptothics} give  $\Pro\left [E_i <  \max_{j \notin \hat{S}} E_{j}\right]= 1 -o(1)$. From the discussion above, we also have $\Pro[S \neq \hat S] \geq \Pro\left [E_i <  \max_{j \notin \hat{S}} E_{j}\right]$, thus implying that the probability of failing tends to $1$.
\end{IEEEproof}

\begin{theorem}[Fisher–Tippett–Gnedenko theorem \cite{welsch1973convergence}]
\label{theo:ftg}
Let $E_1 , E_2, \dots E_n$ be a sequence of independent and identically-distributed Gaussian random variables, distributed as $E_i \sim \mathcal{N}(0,1)$, and $M_{n}= \max\{E_{1},\ldots ,E_{n}\}$. For the pairs of real numbers 
\begin{align*}
&a_{n} = (2\log n)^{-\frac{1}{2}} \\
&b_{n} = (2\log n)^{\frac{1}{2}} - \frac{1}{2} (2\log n)^{-\frac{1}{2}} (\log \log n + \log (4\pi))
\end{align*}
the normalized cumulative distribution function converges as \begin{equation*}
\lim_{n\to \infty} \mathbb{P}\left(\frac{M_{n}-b_{n}}{a_{n}} \leq x\right) = F(x)
\end{equation*} 
where $F(x) = e^{-e^{-x}}$ is the Gumbel cdf.
\end{theorem}

%% file: planted_k_subgraph_appendix.tex
\begin{IEEEproof}[Proof of \cref{theo:subgraph_ML}]
\begin{equation*}
\begin{split}
\Pro_e &= \Pro_{S,E}\left[\hat{S}(E) \ne S\right] \\ 
&= \sum_{S \colon |S| = k} \int dE \ \Pro[S] p(E | S) \mathbbm{1}[\hat{S}(E) \ne S]
\end{split}
\end{equation*}
hence the algorithm the minimizes the error is the one the maximizes $p(E | S) \mathbbm{1}[\hat{S}(E) = S]$ for every clique $S$. 
\begin{equation*}
\begin{split}
&\hat{S}^{MAP}(E) = \hat{S}^{ML}(E) =  \argmax_{S \colon |S| = k} p(E | S) \\
&= \argmax_{S \colon |S| = k} \prod_{l,m \notin S}
\mathcal{N}\left(E_{lm}| 0,\sigma^2\right)
\prod_{i,j \in S}
\mathcal{N}\left(E_{ij}| \mu,\sigma^2\right) \\
&= \argmax_{S \colon |S| = k} \prod_{i,j \in S} \frac{\mathcal{N}\left(E_{ij}| \mu,\sigma^2\right)}{\mathcal{N}\left(E_{ij}| 0,\sigma^2\right)} \prod_{l,m}
\mathcal{N}\left(E_{lm}| 0,\sigma^2\right) \\
&= \argmax_{S \colon |S| = k} \prod_{i,j \in S} exp\left(\frac{E_{ij}}{\sigma^2}\right) = \argmax_{S \colon |S| = k} \sum_{i,j \in S} E_{ij}
\end{split}
\end{equation*}
\end{IEEEproof}

\begin{IEEEproof}[Proof of \cref{theo:k_subgraph_recovery}]
\textbf{Lower bound}
The event of fail for the ML estimator reads:
\begin{align*}
\bigcup_{\substack{\hat{S} \colon |\hat{S}| = k \\  \hat{S} \ne S}} \{W(\hat{S}) > W(S)\}
\end{align*}
Let us now rewrite it distinguishing the contributions given by non-planted solution $\hat S$ with different overlap $m = \hat S \cap S$, with $m$ ranging from $0$ to $k-1$.
\begin{equation}
\label{eq:event_fail}
\bigcup_{\substack{\hat{S} \colon |\hat{S}| = k \\  \hat{S} \ne S}} \{W(\hat{S}) > W(S)\} = \bigcup_{m=0}^{k-1} \bigcup_{\substack{\hat{S} \colon |\hat{S}| = k \\  |\hat{S} \cap S| = m}} \ \{W(\hat{S}) > W(S)\}
\end{equation}
For every such $\hat S$ with intersection (common nodes) $I$ of size $m$ (there are $\binom{k}{m}\binom{N-k}{k-m}$ many) let us build a set of solutions $\mathcal C_{\hat S}$, called an \emph{independent coverage}, with the following property:
\begin{enumerate}
\item $\hat S \in \mathcal C_{\hat S}$.
\item Any two solutions $\hat{S}_1$ and $\hat{S}_2 \in \mathcal{C}_{\hat S}$ share \emph{all and only} the $\binom{m}{2}$ edges inside $I$.
\end{enumerate}
Intuitively, the latter condition says that we can remove all the $\binom{m}{2}$ edges inside $I$ from our consideration, and restrict to the remaining edges which result in independent random variables. 
Consider
\[
W_{REM}(\hat S) := W(\hat S) - W(I)
\]
for any solution $\hat S$ as above (including the planted solution $S$), and note that 
\begin{equation}
    \label{eq:cost-without-intersection}
    W(\hat S) > W(S) \Leftrightarrow W_{REM}(\hat S) > W_{REM}(S) \  .
\end{equation}
It is easy to build such an \emph{independent coverage} grouping all nodes not in $S$ by $k-m$. This gives a coverage with cardinality 
\begin{equation}
M_m := |\mathcal{C}_{\hat{S}}| = \left \lfloor\frac{N-k}{k - m}\right \rfloor
\end{equation}
Note that this cardinality depends only on the size of the overlap $m$ and not on the actual solution $\hat{S}$.
It is easy to see that the union over all the non-planted solution with overlap $m$ can be written as the union over a set of seed solutions $\mathcal{C}_m$ of the independent coverage of the seeds. Formally
\begin{equation}
\{ \hat{S} \colon |\hat{S}| = k , \  |\hat{S} \cap S| = m \} = \bigcup_{\hat{S} \in \mathcal{C}_m} \mathcal{C}_{\hat S}
\end{equation}
Now we can rewrite the event in \cref{eq:event_fail} as 
\begin{equation}
\begin{split}
\bigcup_{\substack{\hat{S} \colon |\hat{S}| = k \\  \hat{S} \ne S}} &\{W(\hat{S}) > W(S)\}  =\\
&= \bigcup_{m=0}^{k-1} \bigcup_{\hat{S}' \in \mathcal{C}_m} \bigcup_{\hat{S} \in \mathcal{C}_{\hat{S}'}} \ \{W(\hat{S}) > W(S)\}
\end{split}
\end{equation}
We can now use the union bound to establish a lower bound on the probability of success of the ML estimator as:
\begin{align*}
1-\Pro_e &= 1-\Pro\left[\bigcup_{m=0}^{k-1} \bigcup_{\hat{S}' \in \mathcal{C}_m} \bigcup_{\hat{S} \in \mathcal{C}_{\hat{S}'}} \ \{W(\hat{S}) > W(S)\} \right] \\ 
&\ge 1-\sum_{m=0}^{k-1} \sum_{\hat{S}' \in \mathcal{C}_m} \Pro \left[ \bigcup_{\hat{S} \in \mathcal{C}_{\hat{S}'}} \ \{W(\hat{S}) > W(S)\}\right] \\
&\ge 1- k \max_{m} \left[ L_m \Pro_e^m \right]
\end{align*}
where 
\[
\Pro_e^m :=  \Pro \left[ \bigcup_{\hat{S} \in \mathcal{C}_{\hat{S}'}} \ \{W(\hat{S}) > W(S)\}\right]
\]
depends only on the overlap $m$ and not on the solution $\hat S'$, and where we bounded the cardinality of $|\mathcal{C}_m|$ as
\[
|\mathcal{C}_m| \le L_m := \frac{\binom{k}{m} \binom{N - k}{k-m}}{M_m}
\]
We can now use the results on the P-REM to upper bound $\Pro_e^m$. Observe that each $\mathcal C_{\hat S'}$ corresponds to a P-REM with $1+M_m$ states and that using \eqref{eq:cost-without-intersection} we can rewrite $\Pro_e^{m}$ as 
\[
\Pro_e^m =  \Pro \left[ \bigcup_{\hat{S} \in \mathcal{C}_{\hat{S}'}} \ \{W_{REM}(\hat{S}) > W_{REM}(S)\}\right]
\]
where each quantity $W_{REM}(\cdot)$ is given by the sum of the $\ell_m$ edge weights (the edges \emph{not} in the common intersection $I$). Therefore 
\begin{equation*}
\begin{split}
E_{S}=W_{REM}(S) \sim \begin{cases}
\mathcal{N}\left(\ell_m \mu,\ell_m \sigma^2\right) \ &for \ \hat S = S \\
\mathcal{N}\left(0,\ell_m\sigma^2\right) \ &for \ \hat S \neq S
\end{cases}
\end{split}
\end{equation*}
and these random variables are \emph{independent} since, 
in $\mathcal C_{\hat S'}$, solutions do not share edges other than those in $I$ (which are removed in the definition of $W_{REM}(\cdot)$).
Hence the probability $\Pro_e^m$ is equal to the probability of error in a P-REM with parameters $(\ell_m \hat \mu \log N, \sqrt{\ell_m} \hat \sigma \sqrt{\frac{\log N}{2}}, M_m)$.
Let us now compute the SNR for this particular P-REM:
\begin{equation*}
\begin{cases}
\begin{aligned}
&\ell_m \hat \mu \log N = \hat \mu_{REM} \log M_m \\
&\sqrt{\ell_m} \hat \sigma \sqrt{\frac{\log N}{2}} = \hat \sigma_{REM} \sqrt{\frac{\log M_m}{2}}
\end{aligned}
\end{cases}
\end{equation*}
for which follow that the SNR reads
\[
\gamma_m = \frac{\hat \mu_{REM}}{\hat \sigma_{REM}} = \frac{\hat \mu}{\hat \sigma} \sqrt{\frac{\ell_m \log N}{\log M_m}} =  \gamma \sqrt{\frac{\ell_m \log N}{\log M_m}}.
\]
We shall prove that, for every P-REM corresponding to the overlap $m$, it holds that 
\begin{equation}
\label{eq:clique:fail:m}
\Pro_e^m = o\left(\frac{1}{k \cdot L_m}\right).
\end{equation}
where the asymptotic notiation is here referred to $N \to +\infty$. Since $k \leq N^\alpha$ and $m \leq k-1$, 
\begin{align*}
M_m = \left\lfloor\frac{N-k}{k-m}\right\rfloor \geq \frac{N-k}{k-m} - 1
&\ge \frac{N-N^{\alpha}}{N^{\alpha}} -1 = \\
& =N^{1-\alpha} -2 \ .
\end{align*}
This implies together with the inequalities $\binom{k}{m} \le 2^k$ and $\binom{N-k}{k-m} \le  e^{k-m} \left(\frac{N-k}{k-m}\right)^{k-m} \le  e^k(N-k)^{k-m}$:
\begin{align*}
k \cdot L_m & := k \cdot \frac{\binom{k}{m} \binom{N - k}{k-m}}{M_m} \\ 
&\le k \cdot \frac{2^k \cdot e^k \cdot (N-k)^{k-m}}{N^{1-\alpha}-2}\\
& = k \cdot N^{k \frac{1+\log2}{\log N}} \cdot  N^{k-m - 1 + \alpha} (1+o(1))\\
& \le N^{k \frac{1+\log2}{\log N}} \cdot  N^{k-m - 1 + 2\alpha} (1+o(1))\\
\end{align*}
where the last inequality follow from $k \le N^{\alpha}$. Now we can use the asymptotics of the probability of fail $\Pro_e^m$ given in \eqref{eq:REM_asymptothics}. In particular, if $\gamma_m > 2 $, \cref{eq:clique:fail:m} holds whenever
\begin{align*}
\frac{1}{M_m^{\frac{\gamma_m^2}{2} -1}} = o\left( \frac{1}{N^{k \frac{1+\log2}{\log N} + k - m -1+ 2\alpha}} \right),
\end{align*}
and since $M_m \geq N^{1-\alpha} -2 = N^{1-\alpha}(1+o(1))$, it is sufficient that
\begin{align}
&\left(\frac{\gamma_m^2}{2}-1\right)(1-\alpha) > \left(1+\frac{1+\log2}{\log N}\right) k - m -1+ 2\alpha \nonumber
\\
&\left(\frac{\gamma_m^2}{2}-1\right) > \frac{ \left(1+ \frac{1+\log2}{\log N} \right) k-m}{1 -\alpha} -1 +\frac{\alpha}{1-\alpha}\nonumber
\\
&\gamma_m > \sqrt{\frac{2}{1-\alpha}\left(\left(1+ \frac{1+\log 2}{\log N} \right) k - m + \alpha\right)} \nonumber 
\\ 
&\gamma > \sqrt{\frac{2}{1-\alpha}\frac{\left(1+ \frac{1+\log 2}{\log N} \right)k - m + \alpha}{\ell_m}\frac{\log M_m}{\log N}}.
\label{eq:clique:muhatcl:condition}
\end{align}
Since $\ell_m = \binom{k}{2} - \binom{m}{2} = \binom{k-m}{2} + m (k-m)$ we have 
\begin{align*}
\frac{k-m}{\ell_m} &= \left[\frac{{\binom{k-m}{2} + m (k-m)}}{k-m}\right]^{-1} \\&= \left[\frac{k - m -1}{2}+m\right]^{-1} \\&= \frac{2}{k+m -1} \leq \frac{2}{k-1} \ .
\end{align*}
Also note that since $\ell_m \geq \ell_{k-1}$,
\[
\frac{k}{\ell_m} \leq \frac{k}{k-1}\leq 2\ .
\]
By plugging these inequalities into \eqref{eq:clique:muhatcl:condition}, and since $M_m \leq N$, we get 
\begin{multline*}
\sqrt{\frac{2}{1-\alpha}\frac{(1+\frac{1+\log 2}{\log N}) k - m + \alpha}{\ell_m}\frac{\log M_m}{\log N}} \leq \\
\le 2\sqrt{\frac{1}{1-\alpha}\left(\frac{1+\log2}{\log N} + \frac{1+\frac{\alpha}{2}}{k-1}\right)}.
\end{multline*}
This implies that, for any $\gamma > 2\sqrt{\frac{1}{1-\alpha}\left(\frac{1+\log2}{\log N} + \frac{1+\frac{\alpha}{2}}{k-1}\right)}$ \cref{eq:clique:muhatcl:condition} holds as desired. 

To conclude the proof, we observe that indeed $\gamma_m >2$:  for $m\leq k-2$ we have $\gamma_m > \sqrt{2 (k-m)}>2$, while for $m = k-1$ we have $\ell_{k-1} = k-1$ hence it follows that $\gamma>2\sqrt{\frac{1}{k-1}}=2 \sqrt{\frac{1}{\ell_{k-1}}}$ implies  $\gamma_{k-1} > 2\sqrt{\frac{\log N}{\log M_{k-1}}}> 2$ since $M_m\leq N$ for all $m$.

\textbf{Upper bound}: Consider $m=k-1$ and a corresponding $\mathcal C_{\hat S}$. Note that for this particular $m$ we have $\ell_{k-1}=k-1$, $M_{k-1} = N -k$, and $\gamma_{k-1} = \gamma \sqrt{\ell_{k-1}\frac{\log N}{\log (N-k)}}$. Hence for any constant $\gamma  < \sqrt{\frac{1}{k-1}}$ and for $N$ large enough, it follow that $\gamma_{k-1} < 1$ and thus, by \eqref{eq:REM_asymptothics},   the probability of failing converges to 1
\[
 \Pro_e^{k-1} = 1 - o(1) .
\]
The theorem follows from the bound $\Pro_e \ge \Pro_e^m$ for any $m = 0,\dots,k-1$.

\textbf{Comment}: Following the proof we can see that both upper and lower bound are established according to the worst case scenario in the given regime. In the case of the lower bound, the worst case is given for the smallest SNR by solutions with no overlap with the planted solution ($m=0$), since those have the highest multiplicity $L_m \approx N^{k-m-1+\alpha}$. In the case of the upper bound, the worst case is given at the highest SNR from the solution with the biggest overlap ($m=k-1$), since those have highest fail probability. The whole problem can in turn be seen as a classical entropy-energy trade-off, where $\gamma$ assumes the role of the inverse-temperature, the entropy is given by the multiplicity and the energy term as the fail probability.
\end{IEEEproof}

%% file: appendix_w_hsbm.tex
\begin{IEEEproof}[Proof of \cref{theo:k_subhypergraph_recovery}]
\textbf{Lower bound} Note that, also in this extension, every solution is fully specified by a subset of $k$ nodes. The proof thus follows the same steps as the case $h=2$ (\Cref{theo:k_subgraph_recovery}), which leads to the sufficient condition for exact recovery \eqref{eq:clique:muhatcl:condition} that reads
\begin{align}
\gamma > \sqrt{\frac{2}{1-\alpha}\frac{\left(1+ \frac{1 + \log 2}{\log N} \right)k - m + \alpha)}{\ell_m}\frac{\log M_m}{\log N}}.
\label{eq:clique-hypergraph:muhatcl:condition}
\end{align}
The only difference is that the quantity $\ell_m$ reads 
\[
\ell_m = \binom{k}{h} - \binom{m}{h}\ .
\]  
In particular, we have
\[
\frac{k - m}{\ell_m} = \frac{k - m}{\binom{k}{h} - \binom{m}{h}} \leq  \frac{k}{\binom{k}{h}}  = \frac{h(h-1)!(k - h)!}{(k-1)!} = \frac{h}{\binom{k-1}{h-1}}\ ,
\] 
where the inequality holds for every $m \leq k-1$ and can be proved as follows. We want to show that
\begin{align*}
\frac{\binom{k}{h}}{k}  \leq \frac{\binom{k}{h} - \binom{m}{h}}{k - m}  \ . \end{align*}
For $m \leq h-1$ this is trivially true because $\binom{m}{h}=0$. Otherwise, for $m \geq h$ we can expand the binomial coefficients and write the inequality above as follows:
\begin{align*}
(k-m)\frac{\binom{k}{h}}{k}  \leq \binom{k}{h} - \binom{m}{h}  
\end{align*}
that is
\begin{multline*}
(k-m)\frac{k(k-1)\cdots(k-h+1)}{k}  \le \\ \le  {k(k-1)\cdots(k-h+1)} 
- m(m-1)\cdots(m-h+1)
\end{multline*}
which is equivalent to
\begin{multline*}
k(k-1)\cdots(k-h+1) - m (k-1)\cdots(k-h+1)  \le \\ \le {k(k-1)\cdots(k-h+1)} - m(m-1)\cdots(m-h+1) 
\end{multline*}
that is
\begin{align*}
 - m (k-1)\cdots(k-h+1)  \leq  - m(m-1)\cdots(m-h+1)  	
\end{align*}
which holds since $m \leq k$. 

\textbf{Upper bound}
Also in this case, the proof is essentially the same as $h=2$. Consider a maximal overlap $m=k-1$ and a corresponding $\mathcal C_{\hat S}$, which gives again $M_m = N -k$. Note that now 
$\ell_{k-1} = \binom{k-1}{h-1}$ since all hyperedges in $\hat S \setminus S$ consist of the single node in $\hat S \setminus S$ and any $h-1$ nodes in the common intersection between $S$ and $\hat S$ (which consists of $m=k-1$ nodes). 
Thus,  for any constant $\gamma  < \sqrt{\frac{1}{ \binom{k-1}{h-1}}}$ and $N$ large enough, we have $\gamma_{k-1} = \gamma \sqrt{\ell_{k-1}\frac{\log N}{\log (N-k)}} < 1$. Hence, the probability of failing converges to 1
\[
 \Pro_e^{k-1} = 1 - o(1) .
\]
The theorem follows from the bound $\Pro_e \ge \Pro_e^m$ for any $m = 0,\dots,k-1$.
\end{IEEEproof}